\documentclass[
 twocolumn,
 amsmath,
 amssymb,
 aps,
 prb,
]{revtex4-2} 

\usepackage{graphicx}
\usepackage{dcolumn}
\usepackage{multirow}
\usepackage{adjustbox}
\usepackage{bm}

\usepackage[utf8x]{inputenc}
\usepackage{pslatex}
\usepackage{textcomp} 
\usepackage{hyperref}
\usepackage[T1]{fontenc} 
\usepackage{lmodern} 
\usepackage{breakurl} 

\usepackage{caption}
\usepackage{subcaption} 

\graphicspath{{./figures/}}

\newcommand{\ybfeal}{YbFe$_4$Al$_8$}

\begin{document}
\begin{sloppypar} 

\title{Electronic structure of YbFe$_4$Al$_8$ antiferromagnet: \\ A combined X-ray photoelectron spectroscopy and first-principles study}

\author{W. Marciniak}
\email [Corresponding author: \\] {wojciech.robe.marciniak@doctorate.put.poznan.pl}

\affiliation{Institute of Molecular Physics, Polish Academy of Sciences, M. Smoluchowskiego 17, 60-179 Poznań, Poland}

\affiliation{Institute of Physics, Faculty of Materials Engineering and Technical Physics, Poznań University of Technology, Piotrowo 3, 61-138 Pozna\'{n}, Poland}

\author{G. Chełkowska}
\author{A. Bajorek}

\affiliation{
Institute of Physics, University of Silesia in Katowice, 75 Pułku Piechoty 1A, 41-500 Chorzów, Poland}

\author{A. Kowalczyk}
\author{A. Szajek}
\author{M. Werwiński}

\affiliation{Institute of Molecular Physics, Polish Academy of Sciences, M. Smoluchowskiego 17, 60-179 Poznań, Poland}

\begin{abstract}
Depending on their chemical composition, Yb compounds often exhibit different valence states.
Here we investigate the valence state of YbFe$_4$Al$_8$ using X-ray photoelectron spectroscopy (XPS) and first-principles calculations.
%
%
The XPS valence band of \ybfeal{} consists of two contributions coming from divalent (Yb$^{2+}$) and trivalent (Yb$^{3+}$) configurations. 
The determined value of the valence at room temperature is 2.81.
Divalent and trivalent contributions are also observed for core-level Yb~4$d$ XPS spectra.
%
%
We study several collinear antiferromagnetic models of \ybfeal{} from the first-principles and for comparison we also consider LuFe$_4$Al$_8$ with a fully filled 4$f$ shell.
We predict that only Fe sublattices of \ybfeal{} carry significant magnetic moments and that the most stable magnetic configuration is AFM-C with antiparallel columns of magnetic moments.
We also present a Mullliken electronic population analysis describing charge transfer both within and between atoms.
In addition, we also study the effect of intra-atomic Coulomb U repulsion term applied for 4$f$ orbitals on Yb valence and Fe magnetic moments.
\end{abstract}

\maketitle

\section{Introduction} 
%
%
Yb-based compounds exhibit a number of interesting properties that include mixed valence, heavy fermion behavior, Kondo effect, magnetic ordering, and superconductivity~\cite{subbarao_new_2013}.
The compound considered in this work, \ybfeal{}, belongs to the ThMn$_{12}$ family of intermetalic compounds with the general formula XT$_4$Al$_8$, where X is the rare-earth or actinide element and T is the transition-metal element~\cite{buschow_note_1976, van_der_kraan_magnetic_1977, suski_chapter_1996}.
Other examples of Yb-based compounds in the XT$_4$Al$_8$ family are YbCr$_4$Al$_8$, YbMn$_4$Al$_8$, and YbCu$_4$Al$_8$~\cite{buschow_note_1976}, while the isostructural configuration with Co has not been confirmed.
In this section we will introduce the issue of valency of \ybfeal{} in comparison with other Yb compounds, we will also discuss the issues related to the complex magnetic configuration of \ybfeal{} and present a model of the crystallographic structure of the considered compound.

\subsection{Valence of Yb-based compounds}
%
%
One motivation for studying YbFe$_4$Al$_8$ is to determine its valence.
Most lanthanides in the metallic state are trivalent, except for Eu and Yb, which are divalent~\cite{strange_understanding_1999, johansson_energy_1979}.
Since Yb has sixteen electrons outside the core, a 4$f$ orbital filled with fourteen electrons is usually stable, and two electrons can participate in the bond.
And indeed, a divalent state is observed for metallic Yb~\cite{lang_study_1981}.
However, the small energy difference observed between the divalent and trivalent states means that Yb in compounds often occurs in the trivalent state or states with valence between two and three.

%
The set of valence values estimated from measurements for a series of Yb compounds takes on a wide range from two to three~\cite{klaasse_systematics_1981}.
%
Examples of valence estimates made from measurements are
2.38 for YbFe$_2$Al$_{10}$~\cite{khuntia_contiguous_2014},
2.68 for Yb$_2$Si$_2$Al~\cite{gannon_intermediate_2018}, and
2.93 for YbRh$_2$Si$_2$~\cite{kummer_intermediate_2011}.
Yb compounds with mean valence are usually classified as fluctuating (mixed) or intermediate valence compounds.
While fluctuating valence implies dynamic transition processes between two valence states, intermediate valence implies the simultaneous presence of two valence states with specific ratios.
It is not uncommon to put the same material into both categories based on similar experimental results, where a well-known example is YbAl$_3$~\cite{buschow_intermediate_1977, kumar_pressure-induced_2008, chatterjee_lifshitz_2017}.
A valence band, similar to the one observed for \ybfeal{}~\cite{tolinski_magnetic_2006}, consisting of contributions from Yb$^{2+}$ and Yb$^{3+}$ 
has been seen previously for both materials classified as valence fluctuating, such as
%
%
YbCu$_2$Si$_2$~\cite{matsunami_combining_2008},
YbAl$_2$~\cite{matsunami_photoemission_2012, matsunami_strongly_2013},
YbB$_{12}$~\cite{rousuli_hard_2017},
Yb$_2$Pt$_6$Ga$_{15}$ and Yb$_2$Pt$_6$Al$_{15}$~\cite{rousuli_photoemission_2017},
as well as for materials of intermediate valence, such as
%
%
%
YbFe$_4$Sb$_{12}$~\cite{anno_electronic_2002},
%
%
YbInCu$_4$~\cite{schmidt_x-ray_2005},
YbNi$_{0.8}$Al$_{4.2}$~\cite{tran_intermediate_2006},
YbRh$_2$Si$_2$~\cite{kummer_intermediate_2011},
Yb$_4$Ga$_{24}$Pt$_9$~\cite{sichevych_intermediate-valence_2017}, and
Yb$_2$Si$_2$Al~\cite{gannon_intermediate_2018}.
%
Since our experimental analysis is based solely on XPS measurements, we cannot unambiguously resolve the membership of the compound under consideration, \ybfeal{}, in the class of fluctuating or intermediate valence materials.
However, previous studies remain consistent that the valence of the considered compound is close to three~\cite{buschow_note_1976,suski_chapter_1996, tolinski_magnetic_2006}.
Suski cites hard-to-find results of Shcherba~$et~al.$ indicating for \ybfeal{} a valence of 3.00(5)~\cite{suski_chapter_1996},
which is consistent with the observed slight positive volume deviation of \ybfeal{} from the trend for a series of mostly trivalent RFe$_4$Al$_{8}$ compounds.
As noted by Kummer~$et~al.$~\cite{kummer_similar_2018} compounds in which the deviation from trivalency is small are among the most interesting, since in this regime there is a transition from a magnetically ordered ground state to a paramagnetic one.

\subsection{Magnetic properties of YbFe$_4$Al$_8$}
%
%
Although the \ybfeal{} compound has been known since at least 1976~\cite{buschow_note_1976}, it attracted the most interest in the first decade of the 21st century~
\cite{
drulis_magnetic_2002,
suski_magnetic_2002,
gaczynski_magnetic_2004,
andrzejewski_unusual_2006,
andrzejewski_negative_2006,
tolinski_magnetic_2006,
szymanski_advantages_2008}.
This was probably due to its misassignment~\cite{gurevich_impedance_2001,drulis_magnetic_2002} to the group of RFe$_4$Al$_8$ superconducting compounds, like for example ScFe$_4$Al$_8$ ($T_c$~=~6~K), YFe$_4$Al$_8$ ($T_c$~=~6~K), and YCr$_4$Al$_8$ ($T_c$~=~4.5~K)~\cite{gurevich_impedance_2001, dmitriev_superconductivity_2003}.
However, a later study denied YbFe$_4$Al$_8$ membership in the superconducting group~\cite{andrzejewski_unusual_2006, andrzejewski_negative_2006}.
%
%
A critical analysis of the presence of a superconductive phase in YbFe$_4$Al$_8$ has led to an explanation of the phenomenon of thermally driven magnetization reversal, sometimes also called negative magnetization~\cite{andrzejewski_unusual_2006, andrzejewski_negative_2006}.
The Néel point of the Fe antiferromagnetic sublattice of \ybfeal{} is about 140~K~\cite{
shcherba_peculiarities_1996}.
As the temperature decreases, a decrease in magnetization in the external field is observed, and a change in the sign of magnetization occurs when the compensation temperature exceeds 34~K~\cite{andrzejewski_unusual_2006, andrzejewski_negative_2006}.
Andrzejewski~${et~al.}$ concluded that the negative magnetization comes from antiferromagnetic interactions between the magnetic moments on Yb and the canted effective magnetic moments on Fe.
The above results for YbFe$_4$Al$_8$ are compared by Andrzejewski~${et~al.}$ with the results for isostructural LuFe$_4$Al$_8$ for which the magnetization observed as a function of temperature is always positive, which the authors relate to the absence of magnetic moments on the Lu sublattice~\cite{andrzejewski_negative_2006}.
However, the above explanation of the negative magnetization in YbFe$_4$Al$_8$ is somewhat controversial due to the assumption of the presence of a magnetic moment on Yb atoms, which has not yet been explicitly confirmed, 
while Suski~$et~al.$ indicated the absence of a magnetic moment on Yb in the considered compound~\cite{suski_magnetic_2002}.
Although the magnetic ordering for the Yb sublattice below 8~K was detected with the $^{170}$Yb Mössbauer effect~\cite{felner_magnetism_1979,suski_magnetic_2002} 
and the effective magnetic moment on Yb ion equal to 4.1~$\mu_B$ was indirectly deduced from the magnetic susceptibility measurements~\cite{shcherba_peculiarities_1996},
Suski~$et~al.$ have shown that at temperatures above 200~K the Yb$_{1-x}$Sc$_x$Fe$_4$Al$_8$ alloys follow the Curie-Weiss law with effective magnetic moments of about 7-8 $\mu_B$/f.u. which do not depend on the composition, indicating that all magnetism in this compound is determined by the Fe sublattice~\cite{suski_magnetic_2002}.
In a subsequent paper, Suski showed that the magnetic susceptibility of YbFe$_4$Al$_8$ is strongly field-dependent, even at temperatures clearly above the antiferromagnetic phase transition, suggesting the existence of a ferromagnetic correlation of unknown origin~\cite{suski_about_2007}.
Suski, therefore, suggests that ferromagnetic clusters of Fe impurities are responsible for the appearance of negative magnetization in YbFe$_4$Al$_8$, which may mimic the presence of a second magnetic sublattice~\cite{suski_about_2007}.
This seems likely, given that about 5\% of the Fe atoms in the YbFe$_4$Al$_8$ sample may be located at 8$i$ or 8$j$ sites, as shown by Mössbauer effect measurements~\cite{szymanski_advantages_2008}.
Gaczyński~$et~al.$ suggest that partial disorder leads to interacting antiferromagnetic and spin-fluctuating subsystems~\cite{gaczynski_magnetic_2004}.

%
The thermally driven reversal magnetization observed in YbFe$_4$Al$_8$ is also related to the phenomenon of negative magnetoresistivity detected at low magnetic fields in RM$_4$Al$_8$ compounds (including YbFe$_4$Al$_8$)~\cite{dmitriev_negative_2008}.
A proposed explanation for this phenomenon is a combination of the Kondo effect with spin-glass state resulting from crystallographic disturbance~\cite{dmitriev_negative_2008}.
The occurrence of a similar effect, which was the inverse magnetocaloric effect observed in Y$_{1-x}$Gd$_x$Co$_2$ alloys, has been associated with antiferromagnetic or cluster glass behavior~\cite{pierunek_normal_2017}.

%
\subsection{Crystal structure of YbFe$_4$Al$_8$}
\begin{figure}[t]
\vspace{5mm}
\includegraphics[clip, width = 0.95 \columnwidth]{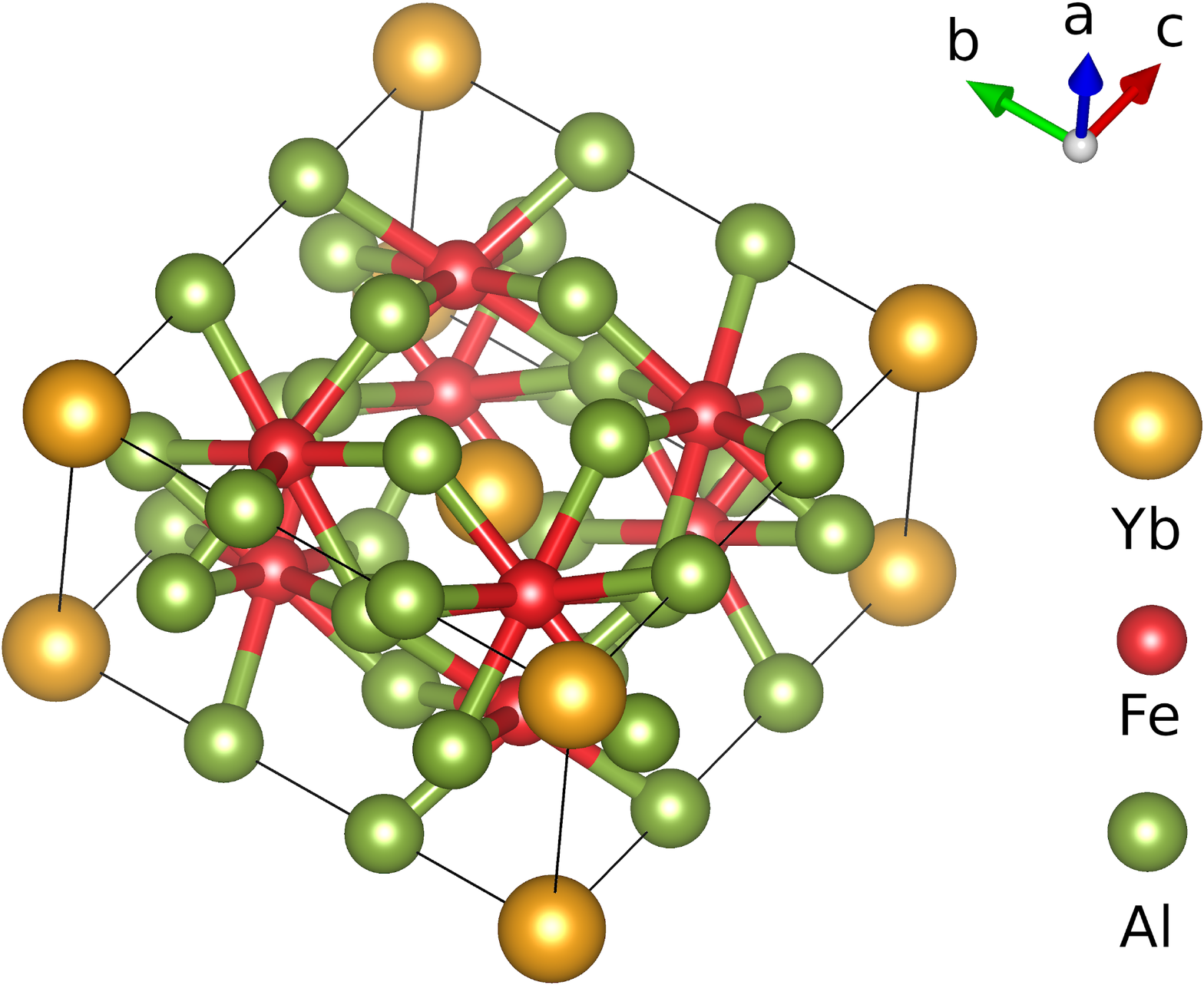}
\caption{\label{fig_struct} 
A model of the crystal structure of the \ybfeal{} compound that crystallizes in a tetragonal structure of the CeMn$_4$Al$_8$ type (space group $I$4/$mmm$, lattice parameters $a$~=~8.712 and $c$~=~5.044~\AA) and is a superstructure of the ThMn$_{12}$ type.
}
\end{figure}
YbFe$_4$Al$_8$ crystallizes in a tetragonal structure of CeMn$_4$Al$_8$ type ($I$4/$mmm$ space group, lattice parameters $a$~=~8.712 and $c$~=~5.044~\AA~\cite{tolinski_magnetic_2006}) that is a superstructure of ThMn$_{12}$ type, see Fig.~\ref{fig_struct}.
Lanthanide contraction is the reason that the unit cell volume of YbFe$_4$Al$_8$ is one of the smallest among the RFe$_4$Al$_8$ compounds~\cite{buschow_note_1976}.
YbFe$_4$Al$_8$ unit cell consists of 26 atoms (two formula units)~\cite{moze_preferential_1990, schobinger-papamantellos_magnetic_1998}.
Yb and Fe atoms occupy the 2$a$ and 8$f$ site, respectively, while Al atoms occupy the 8$i$ and 8$j$ positions.
As shown in the example of YbCu$_4$Ga$_8$, it is also possible to create more complex superstructures in \textit{1-4-8} systems, which require eight non-equivalent atomic positions for a complete description~\cite{subbarao_new_2013}.

\section{\label{exp} Methods}
\subsection{Experimental details}
A polycrystalline YbFe$_4$Al$_8$ sample was obtained by induction melting of stoichiometric amounts of elements in an argon atmosphere.
Details of the preparation are given in earlier work ~\cite{tolinski_magnetic_2006,andrzejewski_unusual_2006,andrzejewski_negative_2006}.
X-ray photoelectron spectra were obtained with an Al-K$_\alpha$ source at room temperature using a PHI 5700/660 Physical Electronics Spectrometer.
The electron energy spectra were analyzed using a hemispherical mirror analyzer with an energy resolution of approximately 0.3~eV.
The Fermi level $E_{\textrm{F}} = 0$ was related to a binding energy of Au~4$f$ at 84~eV.
All emission spectra were measured immediately after breaking up the sample in a vacuum of 10$^{-9}$~Torr.
High vacuum breaking produced clean surfaces free of oxygen and carbon contamination.

%
\subsection{Computational details}
The second part of this paper will present the results of computations performed under density functional theory (DFT).
The \ybfeal{} model was investigated using the full-potential local-orbital scheme (FPLO version 18.00-52)~\cite{koepernik_full-potential_1999,eschrig_2._2004}.
The use of the full potential approach is particularly important for 4$f$-electron compounds, for which the results are strongly dependent on the quality of the potential~\cite{koepernik_full-potential_1999}.
The second important element of our method is the treatment of relativistic effects in the full 4-component formalism.
The use of a fully relativistic method (accounting for spin-orbit coupling) significantly improves the description of 4$f$ electrons characterized by a large spin-orbit coupling.
For the exchange-correlation potential, we chose the generalized gradient approximation (GGA) with the Perdew-Burke-Ernzerhof (PBE) parametrization~\cite{perdew_generalized_1996}.
For elements containing 4$f$ electrons, it is worth investigating how the results might be affected by the inclusion of an additional Coulomb U term of intra-atomic repulsion to the energy functional, described by the GGA~+~U method~\cite{ylvisaker_anisotropy_2009}.
In the present work, we used the fully-localized limit of the LSDA+U (GGA+U) proposed 
by Czy{\.z}yk and Sawatzky~\cite{czyzyk_local-density_1994}, which is also known as the atomic limit
The effect of the value of the parameter U on the obtained results in the range from 0 to 10 eV has been investigated.
Previously, a similar LDA~+~U(4$f$) approach was adopted for YbFe$_4$Sb$_{12}$~\cite{schnelle_itinerant_2005,sichelschmidt_optical_2006}.
Furthermore, our previous experience with modeling alloys containing 3$d$ elements shows that including the LDA+U correction for 3$d$ atoms has very little effect on the position of the 3$d$ band~\cite{morkowski_x-ray_2011,skoryna_xps_2016}, so to simplify the model we have not included this correction.
The calculations were performed on a 20$\times$20$\times$20  k-mesh with a density accuracy of 10$^{-6}$.
The site geometry optimization was performed with a force accuracy of 10$^{-3}$~eV\,\AA{}$^{-1}$.
Details of the geometry optimization will be presented in the next section.
Drawings of the crystalline and antiferromagnetic structures were made using the VESTA code~\cite{momma_vesta_2008}.

%
\section{Results and discussion}
The properties of \ybfeal{}, such as the thermally driven magnetization reversal, magnetism of Yb ions, non-trivial valence and antiferromagnetic configuration motivated us to perform studies that we hope will resolve some of the ambiguities.
In our previous work on YbFe$_4$Al$_8$ we investigated its magnetic and transport properties~\cite{tolinski_magnetic_2006,andrzejewski_unusual_2006,andrzejewski_negative_2006}.
Here we will extend the analysis of X-ray photoelectron spectra presented before~\cite{tolinski_magnetic_2006}.
The analysis of the valence band will be followed by the interpretation of selected core-level spectra. 
In the second part of the work, we will present the results of first-principles calculations focusing on the structural and magnetic properties of the system.

\subsection{X-ray photoelectron spectroscopy \label{xpsexp}}
\subsubsection{Valence band spectrum}
\begin{figure}[t]
\includegraphics[clip, width = 1.0 \columnwidth]{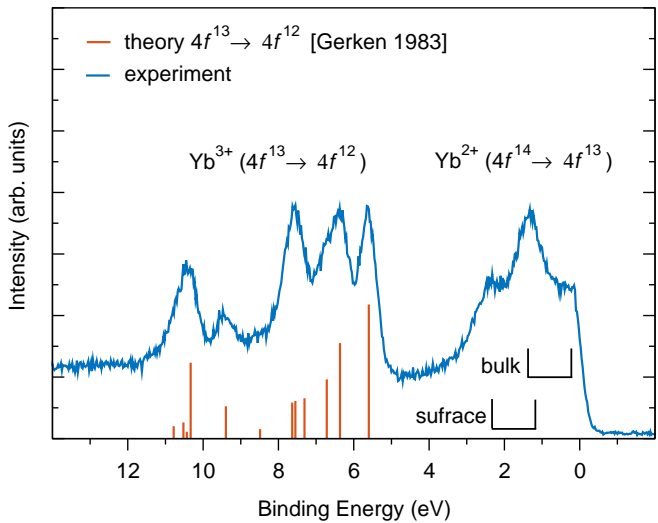}
\caption{
Valence band of YbFe$_4$Al$_8$ measured by X-ray photoelectron spectroscopy at room temperature using Al~K$_{\alpha}$ source with photon energy 1487.6~eV. 
The experimental result is compared with the multiplet structure of the 4$f^{13}$ $\rightarrow$ 4$f^{12}$ transition calculated by Gerken~\cite{gerken_calculated_1983}.
}
	\label{fig-XPS-ybfe4al8-valence}
\end{figure}
%
%
While for divalent metallic Yb we observe a rather simple valence band with a 2F spin-orbit doublet~\cite{lang_study_1981}, Yb compounds with a mean valence, between two and three, reveal a valence band composed of two parts associated with different states before photoemission~\cite{eggenhoffner_advances_1993}.
Such binary valence band spectra have been observed before for YbFe$_4$Sb$_{12}$~\cite{anno_electronic_2002} and
YbInCu$_4$~\cite{schmidt_x-ray_2005}, for example.
As we see in Fig.~\ref{fig-XPS-ybfe4al8-valence}, the valence band of YbFe$_4$Al$_8$
also consists of two contributions, from divalent (Yb$^{2+}$) and trivalent (Yb$^{3+}$) configurations. 
The divalent part (Yb$^{2+}$) is characterized by a doublet derived from the final state 4$f^{13}$, while the trivalent part (Yb$^{3+}$) is characterized by a multiplet derived from the final state 4$f^{12}$.
The distance separating the two parts, which in this case is just under 6~eV, is interpreted as the intra-atomic correlation energy~\cite{eggenhoffner_advances_1993}.

%
However, we see that the divalent part does not resemble a doublet.
This is because XPS measurements using a source with an energy of about 1.5~keV are sensitive to surface effects, resulting in the appearance of a second doublet shifted by about 1~eV toward higher binding energies~\cite{kummer_intermediate_2011}.
Depending on the mutual shift of the doublets, a structure of four or three peaks may appear. 
A structure similar to the triple one seen in Fig.~\ref{fig-XPS-ybfe4al8-valence} has been observed, for example, for Yb$_2$Pt$_6$Al$_{15}$~\cite{rousuli_photoemission_2017}.
A one way to discard the signal from the surface would be to use hard X-rays with photon energy on the order of a few keVs~\cite{rousuli_photoemission_2017}.
Alternatively, also reducing the energy of the source allows to determine the impact of the surface~\cite{kummer_intermediate_2011}.
Since at a binding energy of about 1~eV there is a maximum in the signal coming from the Fe~3$d$ states, it can be suspected that they may significantly affect the valence band in the vicinity of the Fermi level~\cite{tolinski_magnetic_2006}.
Nevertheless, a comparison of photoionization cross-sections for Yb~4$f$ (0.0820) and Fe~3$d$ (0.0022) orbitals contradicts this speculations~\cite{yeh_atomic_1985}.
The almost forty-fold difference in favor of Yb~4$f$ states, combined with their much larger localization leads to the conclusion that the contribution from Fe is not significant in the considered range of a few eV from the Fermi level.
Similarly, also the contributions from Yb~5$d$~6$s$ and Al~3$s$~3$p$ states can be neglected~\cite{eggenhoffner_advances_1993}.
Empirical confirmation of the above predictions is provided by comparing the XPS valence bands of YbFe$_4$Sb$_{12}$ and LaFe$_4$Sb$_{12}$, showing an absolutely dominant contribution from the Yb~4$f$ states~\cite{okane_photoemission_2003}.

%
Now we turn to the trivalent multiplet.
In Fig.~\ref{fig-XPS-ybfe4al8-valence} we compared the experimental result with the multiplet structure of the 4$f^{13}$ $\rightarrow$ 4$f^{12}$ transition calculated by Gerken~\cite{gerken_calculated_1983}.
However, we shifted the transition diagram by 5.6~eV towards higher binding energies.
We determined the value of the 4$f$ level shift relative to the Fermi level from the energy difference of the trivalent and tetravalent metallic Yb calculated by Johansson~\cite{johansson_energy_1979}.
The observed surface contribution in the divalent part of the valence band provokes the question of an analogous contribution in the trivalent part.
Although our result obtained for a photon energy of 1486.6~eV is not conclusive on this question, a series of measurements made for YbRh$_2$Si$_2$ at beam energies of 600, 350, and 115~eV revealed that also the multiplet of trivalent Yb consists of contributions from the bulk and surface~\cite{kummer_intermediate_2011}.
However, the small difference in their mutual position (on the order of 0.2~eV) does not allow us to distinguish them as easily as in the case of the divalent part of the valence band.
In addition to the beam energy, the observed valence band spectrum is also affected by the temperature.
For example, XPS valence bands measured for YbInCu$_4$ at temperatures of 10 and 150~K showed that at a lower temperature the trivalent contribution decreases and the divalent contribution increases, which is naturally due to the decrease in mean valence (of 0.1 in this case)~\cite{schmidt_x-ray_2005}.

%
The measured valence band spectra can be used to determine valence.
For this purpose, the relative intensity $\eta$ is determined from the formula:
\begin{equation}
\eta = \frac{I^{(2+)}}{I^{(2+)}+I^{(3+)}},
\end{equation}
where $I^{(2+)}$ and $I^{(3+)}$ are the integral intensities~\cite{kummer_intermediate_2011}.
The determined value of $\eta$ is affected by the choice of background correction~\cite{schmidt_x-ray_2005} and the difficulty in separating the surface and volume contributions.
Assuming that surface and the volume components of the divalent contribution are equal, we obtain $\eta$ equal to 0.315.
Although the relationship  $v = 3 - \eta$ is usually used to determine the valence ($v$) from the relative intensities in Yb compounds, Kummer~$et~al.$ have shown that a more accurate result can be obtained using the cubic function
\begin{equation}
v(\eta) = 3 - 0.55\eta - 0.01\eta^2 - 0.44\eta^3.
\end{equation}
Thus, for \ybfeal{} at room temperature, the determined valence value is 2.81.
The effect of surface contributions on the result can be minimized in the future by performing the hard x-ray photoemission spectroscopy (HAXPES)
measurements~\cite{rousuli_photoemission_2017}.
Using the example of YbFe$_4$Sb$_{12}$, Okane~$et~al.$ have shown that low-temperature soft-x-ray synchrotron radiation photoemission spectroscopy (SRPES) can also be very helpful in the valence band analysis of Yb compounds~\cite{okane_photoemission_2003}.

\subsubsection{Core level spectra Yb~4d, Fe~2p, and Al~2p}
%
%
\begin{figure}[t]
\includegraphics[clip, width = 0.95 \columnwidth]{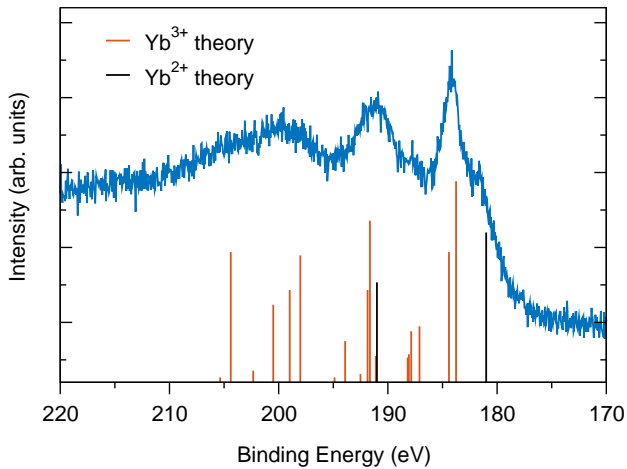}
	\caption{
X-ray photoelectron spectrum of Yb~4$d$ for YbFe$_4$Al$_8$ measured at room temperature using Al~K$_{\alpha}$ source with photon energy 1486.6~eV. 
The experimental result is compared with the multiplet structures of Yb$^{2+}$ and Yb$^{3+}$ computed by Ogasawara~$et~al.$~\cite{ogasawara_lifetime_1994}.	
	}
	\label{fig-XPS-yb-4d}
\end{figure}
The mean valence of YbFe$_4$Al$_8$ also affects its core-level states.
As can be seen in Fig.~\ref{fig-XPS-yb-4d}, the core-level spectrum of the Yb~4$d$ orbital is relatively complex and consists of parts derived from the Yb$^{3+}$ and Yb$^{2+}$ configurations.
The Yb~4$d$ spectrum can be interpreted based on theoretically predicted multiplets calculated for Yb$^{3+}$ and Yb$^{2+}$ by Ogasawara~$et~al.$~\cite{ogasawara_lifetime_1994}.	
The divalent doublet consists of spin-orbit split levels, while the trivalent multiplet arises from more complex 4$d$-4$f$ Coulomb-exchange interactions resulting from incomplete 4$f$ shell occupancy.
The measured spectrum reveals that the contributions from the divalent doublet are rather small, confirming that the valence of the compound is closer to three.
This also agrees with the result deduced for the valence band.
The presented Yb~4$d$ spectrum can also be compared with the results for Yb metal~\cite{hagstrom_electron_1970} and other Yb compounds~\cite{szytula_electronic_2003,schmidt_x-ray_2005,rai_intermediate_2015}.
This comparison confirms the nature of the spectrum similar to the trivalent state of Yb with a smaller divalent contribution.
For mixed-valence compound YbInCu$_4$~\cite{schmidt_x-ray_2005}, it has been observed that the intensity of the 4$d_{5/2}$ divalent peak around 182~eV strongly decreases with decreasing temperature, which was associated with an increase in valence.
Whereas for the mixed-valence alloy (Lu$_{1-x}$Yb$_x$)$_3$Rh$_4$Ge$_{13}$~\cite{rai_intermediate_2015} it has been shown that the peak associated with the Yb$^{3+}$ state located around 185~eV increases significantly with increasing Yb concentration.

%
\begin{figure}[t]
\includegraphics[clip, width = 1.0 \columnwidth]{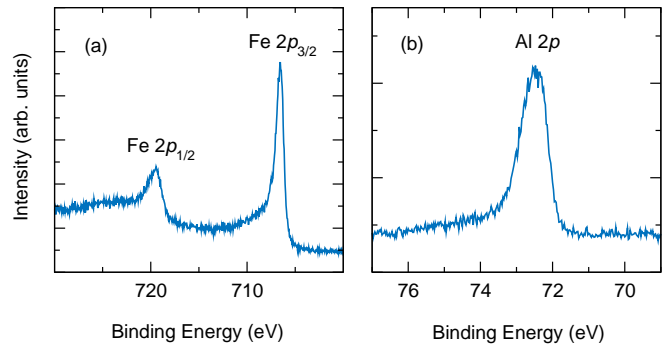}	
	\caption{The XPS spectra of Fe~2$p$ (a) and Al~2$p$ (b) states for YbFe$_4$Al$_8$ measured at room temperature using Al~K$_{\alpha}$ source with photon energy 1486.6~eV.}
	\label{fig-XPS-fe-2p-al-2p}
\end{figure}
The spin-orbit split doublet of asymmetric Fe~2$p$ lines having Doniach-Sunjic shape resembles the corresponding result for pure iron~\cite{graat_simultaneous_1996} and also shows none of the chemical shifts and charge-transfer satellites characteristic of iron oxides~\cite{yamashita_analysis_2008}, see Fig.~\ref{fig-XPS-fe-2p-al-2p}(a).
The determined spin-orbit coupling of Fe~2$p$ doublet (about 13~eV) is similar to the value observed in pure Fe.
%
%
The position of the Al~2$p$ peak (-72.6 eV) observed in Fig.~\ref{fig-XPS-fe-2p-al-2p}(b) is similar to that of pure metal Al (-72.7~eV).
The observed spectral line broadening we ascribe to spin-orbit splitting and differences in level positions for the two non-equivalent Al positions present in the compound.
Relativistic atomic calculations, to be presented later, show that the spin-orbit splitting between the Al~2$p_{1/2}$ and Al~2$p_{3/2}$ states is equal to 0.44~eV.

\subsection{\label{abi}First-principles calculations}
\begin{figure*}[t]
\centering
\includegraphics[clip, width =  \textwidth]{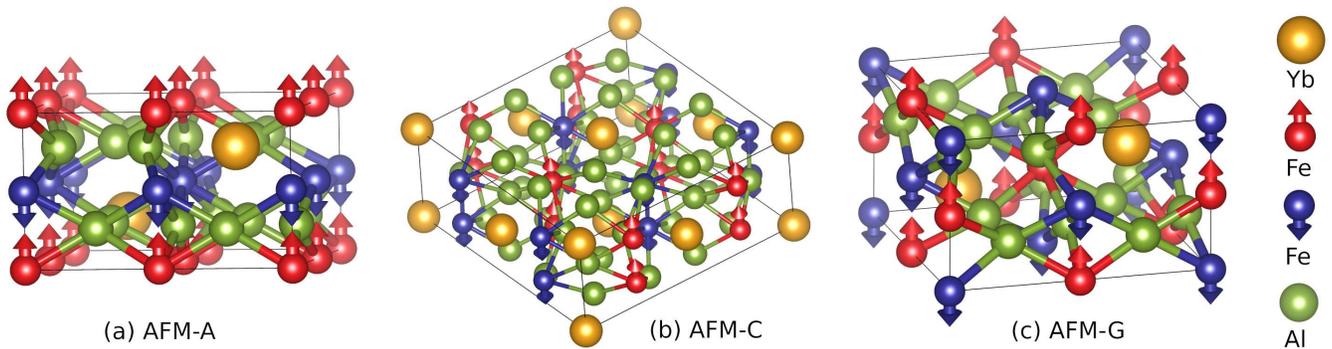}
\caption{\label{fig_afm_stucts} 
Considered antiferromagnetic configurations of spin magnetic moments on Fe sites for \ybfeal{}: AFM-A (a), AFM-C (b), and AFM-G (c).
}
\end{figure*}
\subsubsection{Theoretical models for Yb and \\
 DFT results for XFe$_4$Al$_8$ compounds}

%
Although DFT-based models perform quite well in predicting a range of properties of many rare-earth compounds~\cite{svane_ab_2000}, the single-electron nature of local density approximation (LDA), which does not adequately take correlation into account, severely limits the quality of results obtained for mixed-valence Yb-based compounds~\cite{ylvisaker_charge_2009}.
As a result, the LDA (LDA~+~U) framework yields to a single divalent-type Yb~4$f$ valence band~\cite{jeong_first-principles_2006, yasui_observation_2013}, instead of a complex two-part divalent/trivalent band as observed from spectroscopy.
The successful step toward overcoming this barrier was made in a paper analyzing the effect of pressure on Yb properties, which used LDA-based Quantum Monte Carlo (QMC) methods to determine the valence and to model the valence band of Yb~\cite{ylvisaker_charge_2009}.
An alternative approach to modeling the valence band of Yb compounds is the single-impurity Anderson model (SIAM)~\cite{kummer_intermediate_2011}.
This method, although it leads to a two-part 2+/3+ valence band, is based on the approximation of Yb as an impurity, so it is also not applicable to modeling Yb compounds from first-principles.

%
Earlier DFT results for \ybfeal{} were presented in 2019 by Wang~$et~al.$ in a review paper on RFe$_4$Al$_8$ compounds~\cite{wang_phase_2019}, in which, among other things, the authors calculate lattice parameters of \ybfeal{} (8.703 and 5.049~\AA{}), spin magnetic moment on Fe (1.559~$\mu_B$), elastic constants, phonon spectra, bulk, shear, and Young's modulus.
Their calculations are based on the full potential (FP) projector augmented wave (PAW) approach with Perdew-Burke-Ernzerhof (PBE) approximation. 
In this work, we will improve their model by using a fully relativistic approach and considering an antiferromagnetic configuration.
Previously, the antiferromagnetic configuration of Fe sublattices was also considered in the works of Grechnev~$et~al.$ on RFe$_4$Al$_8$ superconductors (R~=~Sc, Y, Lu)~\cite{grechnev_electronic_2014, logosha_features_2014, zhuravleva_electronic_2016}.

%
\subsubsection{Antiferromagnetic configurations,\\
 geometry optimization, and site preferences}
\begin{table}[!ht]
\caption{\label{tab:optimized_structures} 
Crystallographic data for \ybfeal{} as optimized with the FPLO code. 
The lattice parameters for antiferromagnetic configuration AFM-C are $a$~=~12.306~\AA{} and  $c$~=~5.019~\AA{}, see Fig.~\ref{fig_afm_stucts}(b), 
whereas for basic non-magnetic configuration (NM) lattice parameters are $a$~=~8.702~\AA{} and $c$~=~5.019~\AA{}.
}
\begin{tabular}{cccc|cccc}
\hline \hline
\multicolumn{4}{c}{AFM-C sg. $Fmmm$ (69)} & \multicolumn{4}{|c}{NM sg. $I$4/$mmm$ (139)}   \\
\hline
Yb    	 & 0     	& 0 		& 0		& Yb~2$a$	& 0      	& 0 		& 0 \\ 
Fe$_{up}$& -1/4  	& 0  		& -1/4 	& Fe~8$f$   & 1/4      	& 1/4 		& 1/4 \\
Fe$_{dn}$& 0      	& -1/4 		& -1/4	& Al~8$i$   & 0.3307   	& 0  		& 0 \\
Al$_1$   & 0.1693   & 0.1693 	& 0   	& Al~8$j$   & 0.2784    & 1/2 		& 0 \\ 
Al$_2$   & 0.3896   & -0.1104 	& 0 	&       	&         	&   		&    \\ 
\hline \hline
\end{tabular}
\end{table}

%
Although it is confirmed that there are two magnetic Fe sublattices in the \ybfeal{}, the exact type of antiferromagnetic configuration is not known.
The powder neutron diffraction at 4~K for the isostructural compound CaFe$_4$Al$_8$ has shown that the magnetic moments on the Fe atoms form antiparallel chains oriented along the $c$-direction and slightly canted from the axis~\cite{gvozdetskyi_crystal_2018}, 
which can be described as a canted AFM-C configuration.
Since the FPLO code only allows the calculation of collinear magnetic configurations, we consider for the \ybfeal{} simplified collinear AFM-C model.
As the initial lattice parameters, we take the values 8.714 and 5.026~\AA{} measured for \ybfeal{}~\cite{buschow_note_1976}.
Since the literature does not provide atomic positions for the studied compound, we take initial positions from the isostructural CaFe$_4$Al$_8$ (neutron powder data, $T$~=~4~K)~\cite{gvozdetskyi_crystal_2018}.
As a result of optimizing the lattice parameters and atomic positions, we obtain the AFM-C structure presented in Table~\ref{tab:optimized_structures}.
AFM-C structure can be further reduced to a form with only one Fe sublattice, which results we also present in Table~\ref{tab:optimized_structures}.
We see that the optimized lattice parameters (8.702 and 5.019~\AA{}) are in very good agreement with the experimental results (8.714 and 5.026~\AA{})~\cite{buschow_note_1976}.

In addition, we have prepared two more models of antiferromagnetic configurations, AFM-A and AFM-G, and compared them in terms of the total energy in order to identify the most favorable and hence the most stable structure at the lowest temperature, see Fig.~\ref{fig_afm_stucts}.
The AFM-A configuration proven to be unstable, difficult to converge, and its total energy was considerably higher than for the other two.
In contrast, the AFM-C configuration is 0.7~eV\,f.u.$^{-1}$ more stable than AFM-G, 
so results presented from now on will be based on the AFM-C configuration unless otherwise noted.

%
%
On the sidelines of the work to prepare structural models, we decided to answer the question of the preference of Fe atoms to occupy particular sites in crystal structure.
For this purpose, we prepared three YbFe$_4$Al$_8$ structures in which Fe atoms occupied the 8$f$, 8$i$, or 8$j$ sites.
In agreement with the experiment~\cite{moze_preferential_1990, schobinger-papamantellos_magnetic_1998}, 
our calculations show that Fe atoms prefer to occupy the 8$f$ sites and
that the structures with Fe occupying the 8$j$ and 8$i$ sites are higher in energy by 0.58 and 0.97 eV/Fe atom, respectively.
The calculated values are relatively high~\cite{fu_site_1996}, indicating a rather strong preference for Fe to occupy sites 8$f$.

%
\subsubsection{Effect of on-site Coulomb repulsion U on Yb~4f orbital}
%
%
%
\begin{figure}[t]
\centering
\includegraphics[clip, width =  \columnwidth]{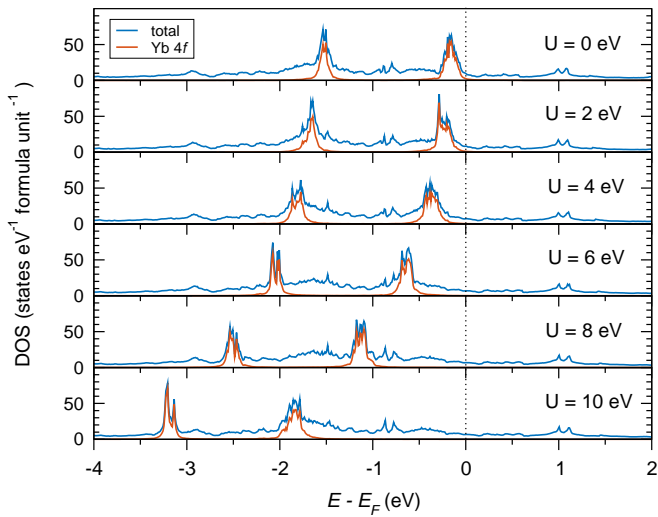}
\caption{\label{fig_dos_vs_u} 
The densities of states of Yb~4$f$ orbitals as a function of on-site repulsion U$_{4f}$ calculated for \ybfeal{} with an antiparallel configuration of magnetic moments on Fe sites (AFM-C). 
Calculations are performed with FPLO18 in a fully relativistic approach, using PBE~+~U, and for the [001] quantization axis.
}
\end{figure}

\begin{figure}[t]
\centering
\includegraphics[clip, width =  \columnwidth]{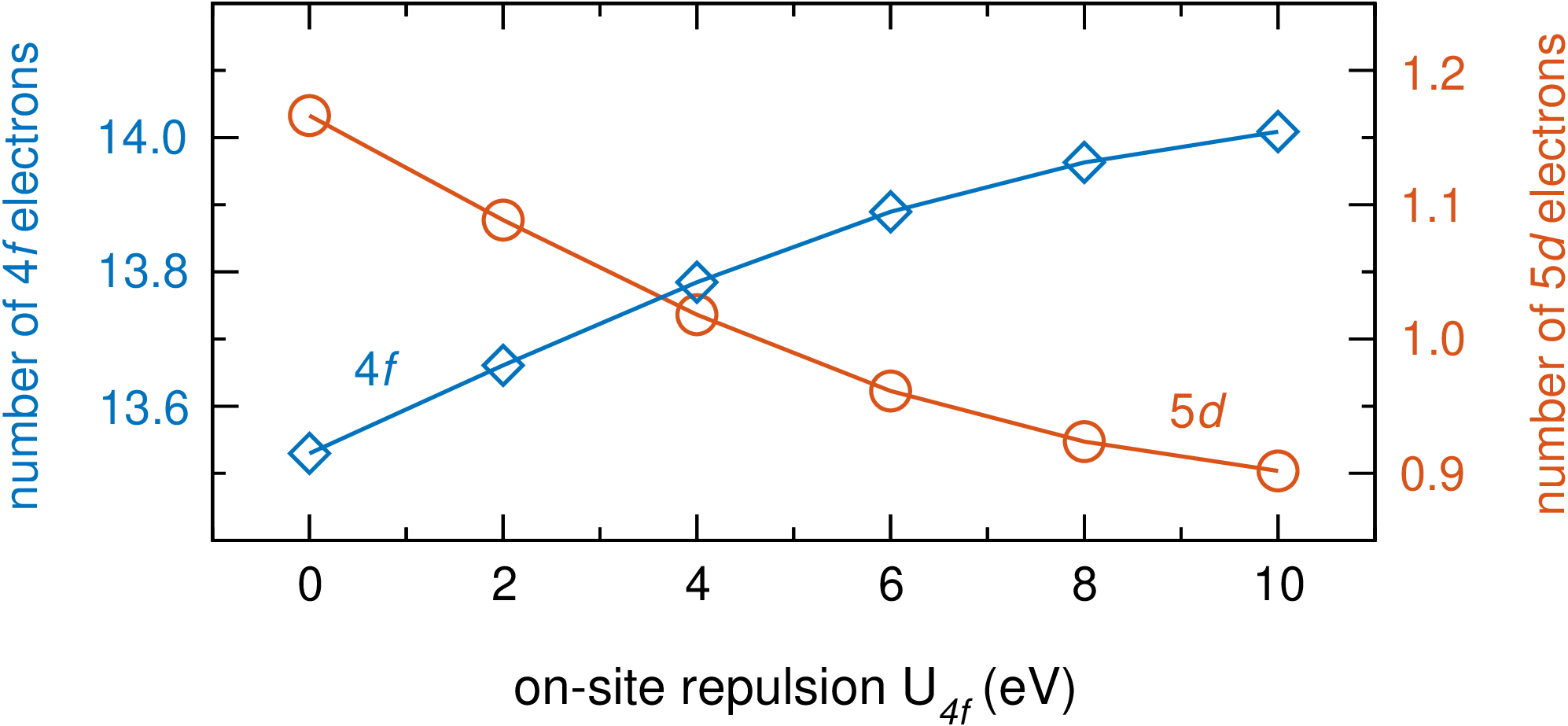}
\caption{\label{fig_occupancy_vs_u} 
Occupancy of Yb~4$f$ and 5$d$ orbitals as a function of on-site repulsion U$_{4f}$ calculated for \ybfeal{} with an antiparallel configuration of magnetic moments on Fe sites (AFM-C). 
Calculations were performed with the FPLO18 in a fully relativistic approach, using PBE~+~U, and for the [001] quantization axis.
}
\end{figure}

LDA+U calculations for YbFe$_4$Sb$_{12}$ showed that the on-site Coulomb repulsion U significantly affects the position of the Yb~4$f$ states in the valence band~\cite{schnelle_itinerant_2005,sichelschmidt_optical_2006}.
In the following, we would like to determine whether the effect that the correction induces is beneficial, against the previously presented valence band measurements. 
For this purpose we examine how the properties of \ybfeal{} change as U applied to 4$f$ orbitals increases from 0 to 10~eV.
%
%
In Fig.~\ref{fig_dos_vs_u} we see that the valence band region of the  densities of states (DOS) is dominated by two spin-orbit split peaks of Yb~4$f$ states. 
As a result of the aforementioned limitations of the GGA/GGA~+~U method, we observe only a single 4$f$ doublet rather than a complex 2+/3+ binary spectrum as in the XPS measurement presented earlier.
We observe, that as U increases the Yb~4$f$ states shift toward lower energies.
Also as U increases, the occupancy of orbital Yb~4$f$ increases, while the occupancy of orbital Yb~5$d$ decreases, see Fig.~\ref{fig_occupancy_vs_u}, which shows the close relationship these two orbitals.
The occupancy of the 4$f$ orbital for \textit{bare} GGA is equal to 13.53, which is closest to the experimental value equal about 13.2 at 300~K.
Given the above, we conclude that the most appropriate approximation for the description of \ybfeal{} is the GGA model without the on-site Coulomb repulsion U.
Therefore, the detailed results presented below will be based specifically on GGA (PBE).

%


\subsubsection{Relativistic atomic energies}

\begin{figure}[t]
\centering
\includegraphics[clip, width = \columnwidth]{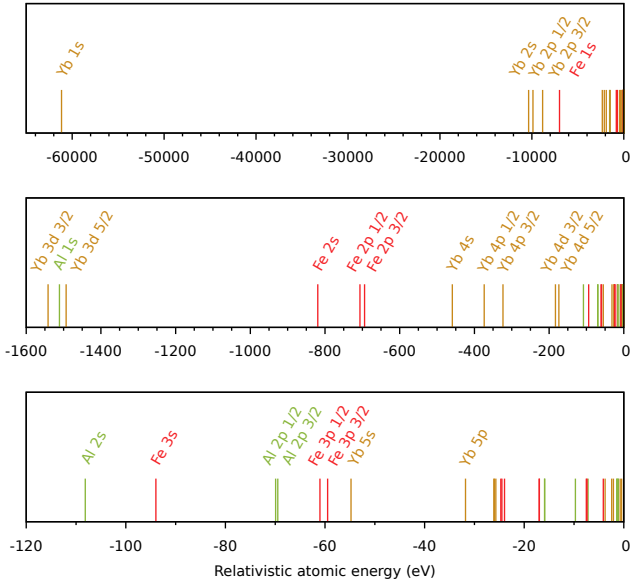}
\caption{\label{fig_ene} 
The relativistic atomic energies for \ybfeal{}
as calculated using FPLO18 in a fully relativistic approach.
}
\end{figure}

Figure~\ref{fig_ene} shows the calculated relativistic atomic energies, which represent the broadest picture of the electronic structure covering all energy levels of \ybfeal{}.
Since these are obtained before the self-consistent cycle, they may change after the calculations converge.
The results are presented in three consecutive ranges and where readability is maintained, energy levels are assigned to individual orbitals.
The levels shown in the top graph do not appear in the results of XPS measurement using a standard X-ray source such as Al-K$_{\alpha}$ with a photon energy of 1486.6~eV.
The lowest level at -61~keV belongs the most strongly bound electron Yb~1$s$.
The middle panel shows the binding energy range typical of an Al-K$_{\alpha}$ source such as the one used in the experimental part of this work, while the bottom panel covers the energy range from -120~eV to the Fermi level.
The calculated levels for Yb~4$d$ (-183.1 and -173.9~eV), Fe~2$p$ (-706.5 and -694.0~eV), and Al~2$p$ (-69.9 and -69.5~eV) are in good agreement with the measured XPS spectra showed before.
The differences are due to 
Coulomb exchange interactions with the valence band levels, 
effect of spin polarization, and in case of Al nonequivalent atomic positions in the unit cell.
Finally, the part of the XPS spectrum, which covers the area of a few eV around the Fermi level can be interpreted more accurately on the self-consistently calculated valence band structure. 
For all-electron methods, like FPLO, accurate positions of the core-electrons energy levels in the range up to about 200 eV of binding energy can be also obtained from self-consistent calculations.

%
\subsubsection{Densities of states}
\begin{figure}[t]
\centering
\includegraphics[clip, width = \columnwidth]{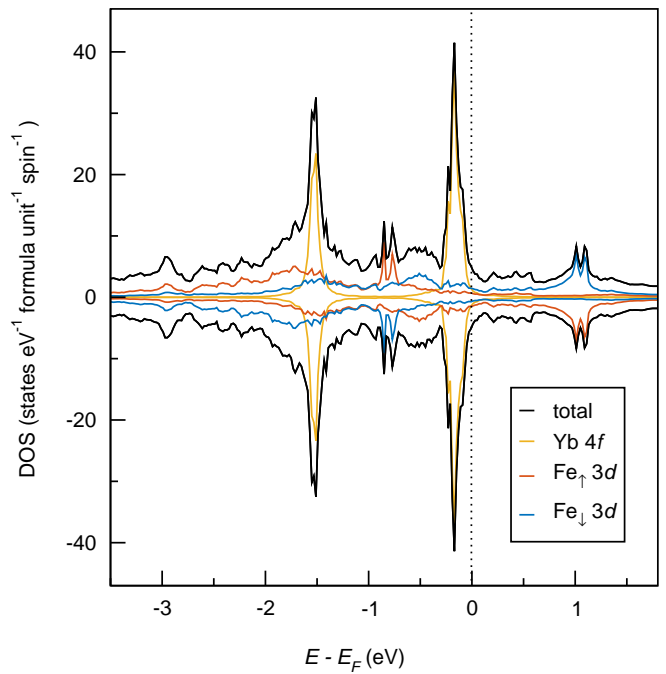}
\caption{\label{fig_dos_afm}
The densities of states (DOS) of \ybfeal{}.
The model is based on the AFM-C configuration of magnetic moments on Fe sublattices (Fe$_{\uparrow}$ and Fe$_{\downarrow}$). 
The total DOS is shown with contributions from selected orbitals (Yb~4$f$ and Fe~3$d$).
Calculations were performed with FPLO18 in a fully-relativistic approach and using PBE functional.
}
\end{figure}
We will start the valence band analysis by presenting the density of states in the range of a few eV around the Fermi level, see Fig.~\ref{fig_dos_afm}.
In this region, the most important contributions come from the Yb~4$f$ and Fe~3$d$ orbitals.
13.53 4$f$ electrons, as determined by Mulliken electron population analysis, uniformly fill the two spin channels. 
However, as a result of spin-orbit splitting, the 4$f$ states split into two narrow bands with energies of about -0.2 and -1.6~eV.
The observed spin-orbit doublet resembles the divalent part of the XPS spectrum, compare Fig.~\ref{fig-XPS-ybfe4al8-valence}.
However, in the PBE picture we do not observe the second trivalent Yb~4$f$ component present in the XPS spectrum below -5~eV.

Since we are considering a model with the antiferromagnetic ordering of moments on Fe atoms,
red and blue colors are used to draw 3$d$ contributions from Fe sublattices with opposite spin polarizations.
As can be seen in Fig.~\ref{fig_dos_afm}, the DOSs of the two Fe sublattices are polarized in opposite directions and their resulting spin magnetic moment is fully compensated.
In the occupied region, the characteristic maxima of the Fe~3$d$ orbitals are located at about -0.8~eV, just between the split Yb~4$f$ bands.
For clarity, the figure does not show contributions from the less occupied valence orbitals, e.g., Yb~5$d$ or Al~3$s$ and 3$p$, a more complete list of which can be found in the next section of this paper on Mulliken analysis.

\subsubsection{Density of states at the Fermi level and\\ 
electronic specific heat coefficient $\gamma$}

%
The density of states at the Fermi level [$\textrm{DOS}(E_{\textrm{F}})$] is 9.4~states\,eV$^{-1}$\,f.u.$^{-1}$ and consists mainly of contributions from Yb~4$f$ (2.5~states\,eV$^{-1}$\,f.u.$^{-1}$) and Fe~3$d$ (3.0~states\,eV$^{-1}$\,f.u.$^{-1}$).
The DOS($E_{\textrm{F}}$) calculated here is relatively close to the values calculated earlier for the anitferromagnetic phases ScFe$_4$Al$_8$, LuFe$_4$Al$_8$, and YFe$_4$Al$_8$ equal to about 14.5~states\,eV$^{-1}$\,cell$^{-1}$~\cite{grechnev_electronic_2014} and to a calculated DOS(E$_{\textrm{F}}$) value for YbFe$_4$Sb$_{12}$ equal to 31~states\,eV$^{-1}$\,f.u.$^{-1}$~\cite{schnelle_itinerant_2005}.

In the Sommerfeld model, the electronic specific heat coefficient $\gamma$ is determined from $\textrm{DOS}(E_{\textrm{F}})$ according to the equation $\gamma = \frac{1}{3} \pi^2 \textrm{k$_{\textrm{B}}^2$} \textrm{DOS}(E_{\textrm{F}})$~\cite{gopal_specific_2012}, where $\textrm{k$_{\textrm{B}}$}$ is Boltzmann constant.
For \ybfeal{} this leads to $\gamma$ equal to 22.2~mJ\,mol$^{-1}$\,K$^{-2}$, which is much lower than experimental $\gamma$ values for isostructural compounds YFe$_4$Al$_8$ (60~mJ\,mol$^{-1}$\,K$^{-2}$) and LuFe$_4$Al$_8$ (75~mJ\,mol$^{-1}$\,K$^{-2}$)~\cite{hagmusa_magnetic_1998, gurevich_impedance_2001}.
However, the underestimation of $\gamma$ is a recognized weakness of the DFT method, resulting from the neglect of spin fluctuations and many-body effects in low-energy excitations~\cite{tanaka_mass_1998}.
Moreover, the $\gamma$ value of 22.2~mJ\,mol$^{-1}$\,K$^{-2}$ calculated for \ybfeal{} is more than twice the value of 9.1~mJ\,mol$^{-1}$\,K$^{-2}$ that we obtained for another compound containing lanthanides and 3$d$-metal (Y$_{0.9}$Ti$_{0.1}$Co$_2$)~\cite{sniadecki_induced_2014}.
The difference is due to the absence of the 4$f$ contribution in the latter case and the slightly higher filling of the 3$d$ shell in Co than in Fe.

%
\subsubsection{Mulliken electronic population analysis}
\begin{table}[t]
\centering
\caption{\label{tab_charge} 
Excess electron number for \ybfeal{} and LuFe$_4$Al$_8$ compounds calculated using FPLO18 in a fully-relativistic approach.
\vspace{2mm}
}
\def\arraystretch{1.5}%
\begin{tabular}{ccccc}
\hline
\hline
formula$\backslash$site	& Yb/Lu	& Fe	& Al1	& Al2 \\
\hline
YbFe$_4$Al$_8$ 			& -0.65	& -0.10	& 0.12	& 0.14 \\
LuFe$_4$Al$_8$  		& -0.88	& -0.08	& 0.14	& 0.17 \\ 
\hline
\hline
\end{tabular}
\end{table}
\begin{table}[t]
\centering
\caption{\label{tab_mulliken} 
The Mulliken electronic population analysis for \ybfeal{} and LuFe$_4$Al$_8$ compounds calculated using FPLO18 in a fully-relativistic approach. Notation of lattice sites of AFM-C structure according to Table~\ref{tab:optimized_structures}.
\vspace{2mm}
}
\def\arraystretch{1.5}%
\begin{tabular}{ccccccc}
\hline
\hline
           		& site  &  6$s$ &  5$d$ &  6$p$ &  4$f$\\
\hline           
\ybfeal{} 		&  Yb   & 0.36  & 1.17 	& 0.32  & 13.53 \\
LuFe$_4$Al$_8$ 	&  Lu   & 0.43  & 1.33  & 0.39  & 13.96 \\
\hline
          		&  site	& 4$s$  &  3$d$ &  4$d$ &  4$p$\\
\hline    	      
\ybfeal{} 		&  Fe 	& 0.65  & 6.73  & 0.11  & 0.42\\
LuFe$_4$Al$_8$ 	&  Fe 	& 0.64  & 6.74  & 0.11  & 0.42 \\
\hline
           		&  site	& 3$s$ &  3$p$ & 3$d$\\
\hline           
\ybfeal{} 		&  Al$_1$ 	& 1.16 & 1.73  & 0.28\\ 
LuFe$_4$Al$_8$ 	&  Al$_1$ 	& 1.15 & 1.75  & 0.28\\ 
\hline
           		&  site	& 3$s$ &  3$p$ & 3$d$\\
\hline           
\ybfeal{} 		&  Al$_2$ 	& 1.12 & 1.75  & 0.30\\
LuFe$_4$Al$_8$ 	&  Al$_2$ 	& 1.12 & 1.77  & 0.31\\
\hline
\hline
\end{tabular}
\end{table}
%

%
%
In YbX$_4$Al$_8$ compounds, the valence of Yb ions is strongly dependent on the transition metal.
For example, the Yb ion is divalent in YbMn$_4$Al$_8$ and possibly in a mixed state in YbCr$_4$Al$_8$ and YbCu$_4$Al$_8$~\cite{felner_magnetism_1979}.
The previous results suggest that Yb ion in YbFe$_4$Al$_8$ is close to trivalent ~\cite{felner_crystal_1978, felner_magnetism_1979}.
However, the valence determined for YbFe$_4$Al$_8$ from the relative intensity of divalent and trivalent peaks in XPS valence band is 2.81 at room temperature, suggesting occupancy of Yb~4$f$ orbitals close to 13.2. 
As shown by the measurements~\cite{schmidt_x-ray_2005}, 
mean valence of the Yb compound may decrease with temperature, suggesting that the Yb~4$f$ occupancy in YbFe$_4$Al$_8$ may be above 13.2 close to 0~K.

Tables~\ref{tab_charge} and \ref{tab_mulliken} show the results of the Mulliken electronic population analysis~\cite{mulliken_electronic_1955}.
The application of Mulliken's approach is possible because the basis of the FPLO code is the method of linear combination of atomic orbitals.
For \ybfeal{}, we observe that the charge taken from the Yb and Fe sites (-0.65 and -0.10) is transferred to the Al sites (+0.12 and +0.14).
The excess electron number for Yb (-0.65) is very different from -2 or -3, values one would expect for a divalent or trivalent configurations.
This is due, among other things, to the fact that first-principles calculations involve a much larger basis than assumed by the conventional valence model (e.g. polarization orbitals Yb~6$s$ and 6$p$) and also allow fractional occupancy of individual orbitals.
Similarly large differences between nominal valence and the calculated occupancy are typically observed in DFT calculations.
For Yb in \ybfeal{} we observe 
fractional occupation of orbital 4$f$ (13.53) and
low occupations of orbitals 5$d$ (1.17) and polarization orbitals 6$s$ and 6$p$ (0.36 and 0.32).
For Fe sites, with regard to the ground state electronic configuration of a neutral Fe atom (3$d^6$ 4$s^2$), we observe an increase in the occupation of the 3$d$ orbital while depopulating the 4$s$ orbital.

%
For comparative analysis, we chose LuFe$_4$Al$_8$, a compound containing a filled 4$f$ shell with a much more stable and simpler to predict electronic structure.
The calculated occupancy of the 4$f$ orbital for LuFe$_4$Al$_8$ is 13.96, which is almost the maximum.
For LuFe$_4$Al$_8$ we observe a similar picture of charge transfer as for YbFe$_4$Al$_8$, but with a significant difference at the position of the 4$f$ element (charge transfer -0.88 for Lu \textit{versus} -0.65 for Yb).
Comparing in more details the occupation of Yb and Lu, we see that at the cost of over-occupation of Yb~4$f$ orbital (above 13.0), the other three Yb orbitals under consideration (5$d$, 6$s$, and 6$p$) are depopulated compared to LuFe$_4$Al$_8$.
Of particular importance is the underoccupancy of the Yb~5$d$ orbital, which leads to a positive deviation of volume from the trend observed for trivalent RFe$_4$Al$_8$ compounds.
While experiments have shown that the type of transition metal affects the valence of Yb in YbM$_4$Al$_8$ compounds~\cite{felner_magnetism_1979}, the change of the 4$f$ element between YbFe$_4$Al$_8$ and LuFe$_4$Al$_8$ compounds shows no significant effect on the occupancy of Fe and Al orbitals. 
However, the greatest changes in electron structure occur on the 4$f$ elements themselves, see Table~\ref{tab_mulliken}.

%
\subsubsection{Spin and orbital magnetic moments}
In the last section we will focus on the detailed analysis of magnetic moments.
For \ybfeal{}, the calculated spin magnetic moments on antiferromagnetically oriented Fe sublattices are equal to 1.523~$\mu_B$ and compensate each other.
The additional orbital contributions occurring on the Fe atoms are 0.037~$\mu_B$/Fe~atom.
Since no magnetic moments have been measured for \ybfeal{} so far, our result can at best be compared with moments experimentally obtained for LuFe$_4$Al$_8$.
Neutron diffraction studies at 1.5 K have shown that for LuFe$_4$Al$_8$ the magnetic moments on the Fe atoms are 1.8(2)~$\mu_B$~\cite{schobinger-papamantellos_magnetic_1998}.
We can see that the value of the total magnetic moment (spin plus orbital) on the Fe atom for \ybfeal{} equal to about 1.6~$\mu_B$ 
is within the range of accuracy of the measurement for LuFe$_4$Al$_8$~\cite{schobinger-papamantellos_magnetic_1998}.
The calculated spin magnetic moments on Fe in \ybfeal{} (1.523~$\mu_B$) are significantly lower than the corresponding values for bcc Fe (2.16~$\mu_{\mathrm{B}}$) calculated under the same theoretical model, as well as compared to the experimental value for bcc Fe (1.98~$\mu_{\mathrm{B}}$~\cite{chen_experimental_1995}).
The calculated orbital magnetic moments on Fe in \ybfeal{} (0.037~$\mu_{\mathrm{B}}$/Fe atom) and in bcc Fe (0.045~$\mu_{\mathrm{B}}$/Fe atom) are comparable.
However, both of these values are reduced from the experimental value for bcc Fe (0.086~$\mu_{\mathrm{B}}$)~\cite{chen_experimental_1995}.
Underestimation of the orbital magnetic moment for transition metals is considered as one of the weaknesses of GGA and LDA.
For \ybfeal{} we do not observe any magnetic moments on the Yb and Al sites.
Since the calculations based on the AFM-C model with a single Yb position cannot account for a possible antiferromagnetic ordering on Yb atoms, we prepared an alternative structural model with two non-equivalent Yb positions in addition to two non-equivalent Fe positions.
The calculations for this model, however, led to identical results as presented for a simpler model with a single Yb position, showing no magnetic moments on the Yb atoms.
However, it is worth noting the limitations of our DFT model, which lead to a description of the Yb~4$f$ valence states that is inconsistent with the results observed in XPS and thus can affect the obtained magnetic state of Yb ions.
In conclusion, although the results of our calculations indicate the absence of a magnetic moment on Yb ions, calculations within a more advanced theoretical model than the one used in this work are needed to unambiguously resolve the question of the magnetic state of Yb ions in \ybfeal{}.

\section{Summary and conclusions}
We have investigated the valence state of YbFe$_4$Al$_8$ using X-ray photoelectron spectroscopy and first-principles calculations.
%
%
We interpreted XPS measurements of the valence band and selected core-levels based on previously predicted theoretical energy level multiplets.
We have  identified that the XPS valence band of \ybfeal{} consists of two contributions coming from divalent (Yb$^{2+}$) and trivalent (Yb$^{3+}$) configuration. 
The YbFe$_4$Al$_8$ valence is determined from the relative intensity of divalent and trivalent peaks is 2.81 at room temperature.

%
The second part of the paper consisted of first-principles calculations performed using a full-potential local-orbital scheme (FPLO).
Since atomic positions of \ybfeal{} are not available in the literature, we performed full structure optimization and presented the resultant complete structural model.
Due to the antiferromagnetic arrangement of magnetic moments revealed in the \ybfeal{} compound, we considered several collinear antiferromagnetic configurations (AFM-A, AFM-C, and AFM-G), of which AFM-C was found to be the most stable.
The AFM-C configuration is characterized by the arrangement of magnetic moments on Fe in the form of antiparallel chains.
The calculations predicted no magnetic moments at the Yb and Al sites and showed that the antiparallel moments at the Fe chains completely compensate.
Mulliken analysis showed that the charge taken from the Yb and Fe sites is transferred to Al and that the fractional occupation of the 4$f$ orbital is about 13.5 (at 0~K), although the valence deduced from the XPS spectra suggests an occupation closer to 13.2 (at 300~K).
The calculated valence band densities of states presenting a spin-orbit split Yb~4$f$ doublet stand in opposition to the complex XPS spectrum consisting of a divalent doublet and a trivalent multiplet of Yb 4$f$ states.
In order to more accurately describe the valence band and resolve the magnetism of Yb ions, calculations beyond the GGA must be performed.

\section*{Acknowledgments}
We acknowledge the financial support of the National Science Centre Poland under the decision DEC-2018/30/E/ST3/00267.
Part of the computations was performed on the resources provided by the Pozna{\'n} Supercomputing and Networking Center (PSNC).
We thank Paweł Leśniak and Daniel Depcik for compiling the scientific software and administration of the computing cluster at 
the Institute of Molecular Physics, Polish Academy of Sciences.
We thank Justyna Rychły and Justyn Snarski-Adamski for reading the manuscript and useful comments.
\end{sloppypar}

\bibliography{ybfe4al8}   

\end{document}